\documentclass[useAMS,usenatbib]{mn2e}
\usepackage{footnote}
\usepackage{graphicx}
\usepackage{amsmath}
\usepackage{natbib}
\usepackage{array}
\usepackage{color}
\usepackage{url}
\usepackage{multirow}
\definecolor{nc}{rgb}{0,0,0}
\def\changed    {\color{nc} }

\definecolor{ncc}{rgb}{0,0,0}
\def\changedbds    {\color{ncc} }

\definecolor{srv}{rgb}{0,0,0}
\def\secondchange    {\color{srv} }

\definecolor{newref}{rgb}{0,0,0}
\def\newref    {\color{newref} }

\definecolor{R4}{rgb}{0,0,0}
\def\Rfour    {\color{R4} }

\definecolor{R5}{rgb}{0,0,0}
\def\Rfive    {\color{R5} }

\def\starpy ~{\textsc{starpy}}

\begin{document}

\title[Quenching Histories of AGN Host Galaxies]{Galaxy Zoo: Evidence for rapid, recent quenching {\Rfive within} a population of AGN host galaxies}
\author[Smethurst et al. 2015]{R. ~J. ~Smethurst,$^{1}$ C. ~J. ~Lintott,$^{1}$ B. ~D. ~Simmons,$^{1, 2,}$\footnotemark[1] K. ~Schawinski,$^{3}$ \newauthor S. ~P. ~Bamford,$^{4}$  C. ~N. ~Cardamone,$^{5}$ S. ~J. ~Kruk,$^{1}$ K. ~L. ~Masters,$^{6}$ \newauthor C. ~M. ~Urry,$^{7}$  K. ~W. ~Willett,$^{8}$ O. ~I. ~Wong$^{9}$ \footnotemark[2]
\\ $^1$ Oxford Astrophysics, Department of Physics, University of Oxford, Denys Wilkinson Building, Keble Road, Oxford, OX1 3RH, UK 
\\ $^{2}$ Center for Astrophysics and Space Sciences (CASS), Department of Physics, University of California, San Diego, CA 92093, USA
\\ $^3$ Institute for Astronomy, Department of Physics, ETH Z\"urich, Wolfgang-Pauli Strasse 27, CH-8093 Z\"urich, Switzerland
\\ $^4$ School of Physics and Astronomy, The University of Nottingham, University Park, Nottingham, NG7 2RD, UK
\\ $^5$ Math \& Science Department, Wheelock College, 200 The Riverway, Boston, MA 02215, USA
\\ $^6$ Institute of Cosmology and Gravitation, University of Portsmouth, Dennis Sciama Building, Barnaby Road, Portsmouth, PO1 3FX, UK 
\\ $^7$ Department of Physics and Yale Center for Astronomy and Astrophysics, Yale University, PO Box 208121, New Haven, CT 06520-8121, USA
\\ $^8$ School of Physics and Astronomy, University of Minnesota, 116 Church St SE, Minneapolis, MN 55455, USA
\\ $^9$ International Centre for Radio Astronomy Research, UWA, 35 Stirling Highway, Crawley, WA 6009, Australia
\\
\\Accepted 2016 August 31.  Received 2016 July 28; in original form 2015 August 4
}

\maketitle

\begin{abstract}

We present a population study of the star formation history of $1244$ Type 2 AGN host galaxies, compared to $6107$ inactive galaxies. A Bayesian method is used to determine individual galaxy star formation histories, which are then collated to visualise the distribution for {\Rfive quenching and quenched galaxies within} each population. We find evidence for {\Rfive some of} the Type 2 AGN host galaxies having undergone a rapid drop in their star formation rate within the last 2 Gyr. AGN feedback is therefore important at least for this population of galaxies.
This result is not seen for the {\Rfive quenching and quenched}  inactive galaxies whose star formation histories are dominated by the effects of downsizing at earlier epochs, a secondary effect for the AGN host galaxies. We show that  histories of rapid quenching cannot account fully for the quenching of all the star formation in a galaxy's lifetime across the population of {\Rfive quenched} AGN host galaxies, and that histories of slower quenching, attributed to secular (non-violent) evolution, are also key in their evolution.
This is in agreement with recent results showing both merger-driven and non-merger processes are contributing to the co-evolution of galaxies and supermassive black holes. The availability of gas in the reservoirs of a galaxy, and its ability to be replenished, appear to be the key drivers behind this co-evolution.
\end{abstract}

\footnotetext[1]{Einstein Fellow}
\footnotetext[2]{This investigation has been made possible by the participation of over 350,000 users in the Galaxy Zoo project. Their contributions are acknowledged at \url{http://authors.galaxyzoo.org}}

\section{Introduction}

The nature of the observed co-evolution of galaxies and their central supermassive black holes \citep{Mag98, MH03, HR04} and the effects of AGN feedback on galaxies are two of the most important open issues in galaxy evolution. AGN feedback was first suggested as a mechanism for regulating star formation in simulations \citep{SR98, Croton06, Bower06, Somer08} and indirect evidence has been observed for both positive and negative feedback in various systems (see the comprehensive review from \citealt{Fab06}). 

The strongest observational evidence for AGN feedback in a population is that the largest fraction of AGN are found in the green valley \citep{CB08, Hickox09, Sch2010}, suggesting some link between AGN activity and  the process of quenching which moves a galaxy from the blue cloud to the red sequence. However, concrete statistical evidence for the effect of AGN feedback on the host galaxy population has so far been elusive.

\begin{figure*}
\includegraphics[width=\textwidth]{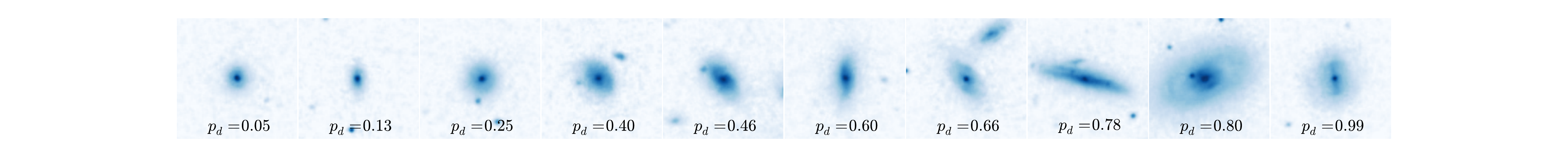}
\caption{Randomly selected SDSS \emph{gri} composite images from the sample of $1,244$ Type 2 AGN in a redshift range $0.04 < z < 0.05$.  The galaxies are ordered from least to most featured according to their debiased `disc or featured' vote fraction, $p_d$ (see \citealt{GZ2}). The scale for each image is $0.099~\rm{arcsec/pixel}$.}
\label{mosaic}
\end{figure*}

Here we present a large observational population study of the {\Rfive star formation histories (SFH)} of Type 2 AGN host galaxies. 
We use a new Bayesian method \citep{Sme2015} to effectively determine the most probable SFH of a galaxy, modelled with two parameters, time of quenching, $t_q$, and exponential rate, $\tau$, given the observed near ultra-violet (NUV) and optical colours. {\changed This builds on the work of \citet{Martin07} and \citet{Sch2014}, but improves significantly on previous techniques. We} aim to determine the following: (i) Are galaxies currently hosting an AGN undergoing quenching? (ii) If so, when and at what rate does this quenching occur? (iii) Is this quenching occurring at different times and rates compared to a control sample of inactive galaxies?

The zero points of all magnitudes are in the AB system. Where necessary, we adopt the WMAP Seven-Year Cosmology (Jarosik et al. 2011) with $(\Omega_m , ~\Omega_\Lambda , ~h) = (0.26, 0.73, 0.71)$.

\section{Data \& Methods}

\subsection{Data Sources}\label{datasource}
In this investigation we use visual classifications of galaxy morphologies from the Galaxy Zoo 2\footnote{\url{http://zoo2.galaxyzoo.org/}} (GZ2) citizen science project \citep{GZ2}, which obtains multiple independent classifications for each optical image. The full question tree for an image is shown in Figure~1 of \citeauthor{GZ2}  The GZ2 project used $304, 022$ images from the Sloan Digital Sky Survey Data Release 7 (SDSS; \citealt{York2000, Abazajian09}) all classified by \emph{at least} 17 independent users, with a mean number of classifications of $\sim42$.

Further to this, we required NUV photometry from the GALEX survey \citep{Martin05}, within which $\sim42\%$ of the GZ2 sample was observed, giving $126, 316$ galaxies total ($0.01 < z < 0.25$). This will be referred to as the \textsc{gz2-galex} sample. The completeness of this sample ($-22 < M_u < -15$) is shown in Figure~2 of \cite{Sme2015}. 

{\changed Observed fluxes are corrected for galactic extinction \citep{Oh11} by applying the \citet*{Cardelli89} law. We also adopt $k$-corrections to $z = 0.0$ and obtain absolute magnitudes from the NYU-VAGC \citep{Blanton05, Pad08, BR07}.}

\subsection{AGN Sample}\label{agnsample}

\begin{figure*}
\includegraphics[width=0.94\textwidth]{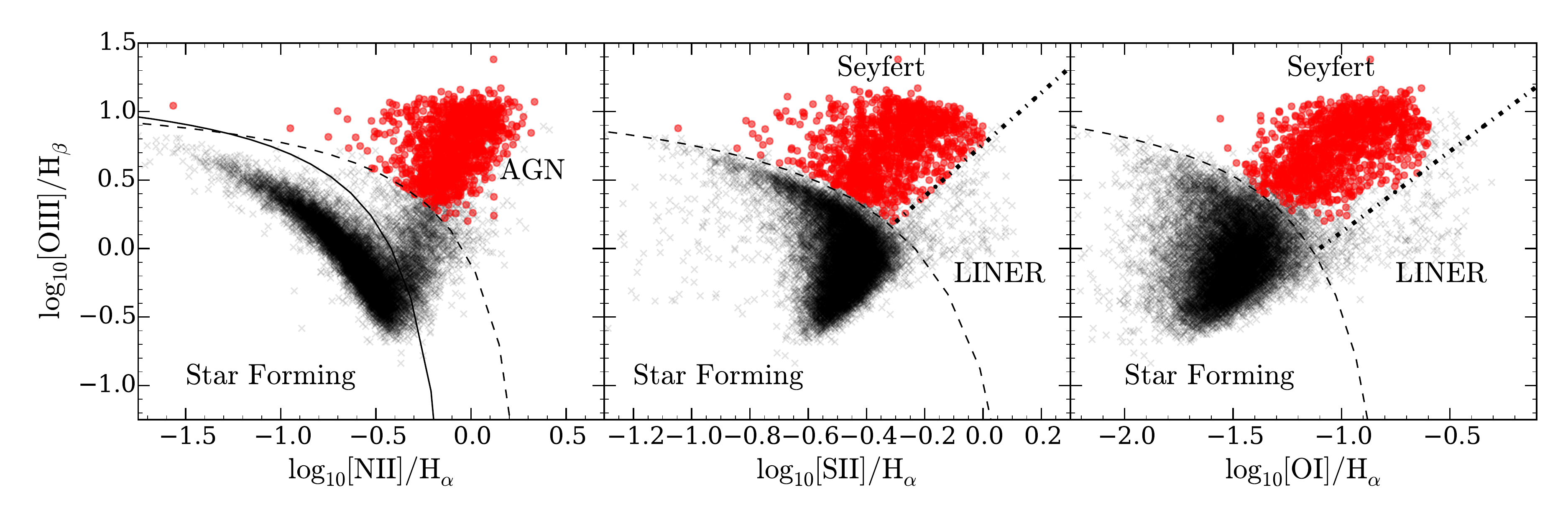}
\caption{BPT diagrams for galaxies in the \textsc{gz2-galex} sample (black crosses) with S/N $> 3$ for each emission line. Inequalities defined in: \protect\cite{Kew01} to separate SF galaxies from AGN (dashed lines), \protect\cite{Kauff03b} to separate SF from composite SF-AGN galaxies (solid line) and \protect\cite{Kew06} to separate LINERS and Seyferts (dotted lines). Galaxies are included in the \textsc{agn-host} sample (red circles) if they satisfy all the inequalities to be classified as Seyferts. LINERs are excluded for purity.}
\label{bpt}
\end{figure*}

We selected Type 2 AGN using a BPT diagram \citep{bpt81} using line and continuum strengths for [OIII], [NII], [SII] and [OII] obtained from the MPA-JHU catalogue \citep{Kauff03, Brinch04} for galaxies in the \textsc{gz2-galex} sample. We then required the S/N $> 3$ for each emission line as in \cite{Sch2010}. Those galaxies which satisfied all of the inequalities defined in \cite{Kew01} and \cite{Kauff03b} were selected as Type 2 AGN, giving $1,299$ host galaxies ($\sim10\%$ of the \textsc{gz2-galex} sample). \cite{Sarzi10, RB12} and \cite{Singh13} have all demonstrated that LINERs are not primarily powered by AGN, therefore for purity, we excluded these galaxies from the sample using the definition from \cite{Kew06} ($55$ galaxies total) with no change to the results. These $1,244$ galaxies will be referred to as the \textsc{agn-host} sample. 

{\changed We refrain from using Type 1 AGN due to concerns about contamination of the SFH analysis from potentially strong NUV emission by unobscured nuclei. The obscuration of Type 2 AGN is highly efficient, considerably more so in the NUV than the optical \citep{Simmons11}; residual NUV flux from a Type 2 AGN can be neglected in comparison to that of the galaxy. We also investigated the possibility of contamination of optical galaxy colours by residual AGN emission, finding that subtracting measured nuclear magnitudes (SDSS {\tt psfMag}) produces a negligible change in host galaxy colour ($\Delta(u-r) \sim 0.09$)}. We therefore use the uncorrected colours to avoid unnecessary complexity and minimise the propagation of uncertainty from the colours through to the SFHs. {\changed However, we note that including these corrected colours does not change our results.}  

{\secondchange We note also that galaxy colours were not corrected for intrinsic dust attenuation. This is of particular consequence for disc galaxies, where attenuation increases with increasing inclination. \cite{Buat05} found the median value of the attenuation in the GALEX NUV passband to be $\sim 1$ mag. Similarly \cite{Masters10} found a total extinction from face-on to edge-on spirals of 0.7 and 0.5 mag for the SDSS $u$ and $r$ passbands and show spirals with $\log(a/b) > 0.7$ have signs of significant dust attenuation. For the \textsc{agn-host} (\textsc{inactive}) sample we find $23\%$ ($25\%$) of discs (with $p_d > 0.5$) have $\log(a/b) > 0.7$, therefore we must be aware of possible biases in our results due to dust. 

From the findings of \cite{Masters10} and \cite{Buat05} above, we estimate the extinction to be $u-r \sim 0.2$ mag and $NUV-u \sim 0.3$ mag, therefore the average change in the SFH parameters across a range of input colours $0 ~<~u-r~<~4$ and $-1~<~NUV-u~<~5$,  are $\Delta~t_q~=~0.985$~Gyr, $\Delta~\tau~=~1.571$~Gyr. This change therefore causes the SFH parameters derived to move towards earlier times and faster quenching rates. Results should be viewed with the caveat, particularly for higher mass, disc galaxies, that earlier values of $t_q$ and more rapid values of $\tau$ may be inferred by \textsc{starpy}. However, we note that (i) applying these average corrections across each sample population does not change our main conclusions, {\newref (ii) that results are consistent if the population of edge-on and face-on galaxies are compared and {\Rfour (iii) that results do not change if only face-on galaxies are used in the investigation}, strongly suggesting that internal galactic extinction does not systematically bias our results.}}

SDSS images for 10 randomly selected galaxies from the \textsc{agn-host} sample are shown in Figure~\ref{mosaic}; Figure~\ref{bpt} shows the entire \textsc{agn-host} sample and the matched \textsc{gz2-galex} galaxies on a BPT diagram.  For the \textsc{agn-host} sample the mean $\log (L[OIII] ~[\rm{erg~s^{-1}}]) \sim 41.3$ and median $\log (L[OIII] ~[\rm{erg~s^{-1}}]) \sim 41.0$, with a range of $\log (L[OIII] ~[\rm{erg~s^{-1}}])$ luminosities of $39.4-43.0$.

We constructed a sample of inactive galaxies by removing from the \textsc{gz2-galex} sample {\changed all galaxies with line strengths indicative of potential AGN activity \citep*{Kauff03b}}, as well as sources identified as Type 1 AGN by the presence of broad emission lines \citep{Oh15}. {\changed We select mass- and morphology-matched inactive samples by identifying between 1 and 5 inactive galaxies for each \textsc{agn-host} galaxy with the same stellar mass (to within $\pm5\%$) and GZ2 `smooth' and `disc' vote fractions (to within $\pm 0.1$); this selects $6107$ galaxies.} We refer to this sample as the \textsc{inactive} sample. A Kolmogorov-Smirnov test revealed the redshift distributions of the \textsc{inactive} and \textsc{agn-host} samples are statistically indistinguishable ($D \sim 0.16$, $p \sim 0.88$).

{\newref We show the \textsc{agn-host} and \textsc{inactive}  samples on both an optical colour-magnitude diagram and in the SFR-stellar mass plane in Figure~\ref{cmdsfms} in comparison to the distribution of SDSS DR7 galaxies. SFRs and stellar masses are obtained from the MPA JHU catalog, where available, which follow the prescriptions outlined in \cite{Brinch04} and \cite{Salim07} for calculating the total aperture corrected galaxy SFR in the presence of an AGN. 

We note that the majority of the \textsc{agn-host} sample would be defined as residing in the blue cloud ($73\%$) on the optical colour-magnitude diagram despite the fact that a {\Rfive significant proportion} of the sample ($47\%$) lie more than $1\sigma$ ($0.3$ $\rm{dex}$) below the star forming ``main sequence'' (fit to the MPA-JHU catalog of SDSS DR7 of \citealt{Kauff03, Brinch04}, see Figure \ref{cmdsfms}). 

{\newref \cite{Ko13} show that in a sample of quiescent red-sequence galaxies without $\mathrm{H}\alpha$ emission, $26\%$ show NUV excess emission and that the fraction with recent star formation is $39\%$. This is more clearly visible in Figure \ref{cmdsfms}b, where a substantial fraction of both the \textsc{agn-host} and \textsc{inactive} samples, all of the sources in which have detected NUV emission, nevertheless lie more than $1\sigma$ below the star-formation sequence.} {\Rfive We do not make a cut on either the \textsc{agn-host} or \textsc{inactive} samples for star formation rate. The SFH of the entire samples are fitted, however we describe in Section \ref{starpy} how our method accounts for those galaxies not appropriately fit by a quenching model and down-weights their contribution to the final results.}

\begin{figure}
\centering
\includegraphics[width=0.4\textwidth]{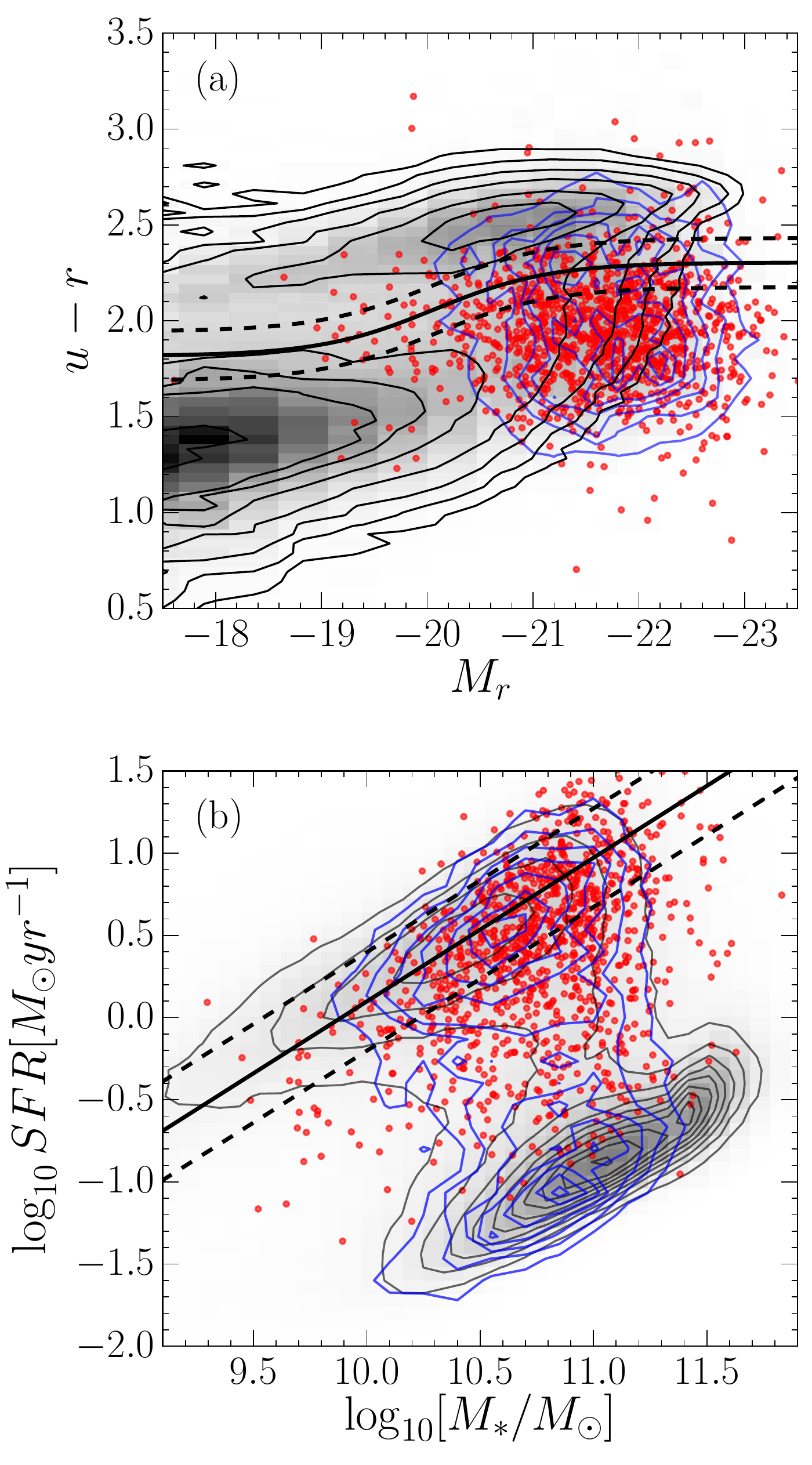}
\caption{{\newref (a) Optical colour-magnitude diagram showing the SDSS DR7 (grey filled contours), the \textsc{agn-host} sample (red circles) and \textsc{inactive} sample (blue contours). The definition of the green valley from \citet{Baldry06} (solid line) with $\pm 1\sigma$ (dashed lines) is shown. (b) SFR-stellar mass diagram showing the MPA-JHU measurements of SFR and $M_*$ of SDSS DR7 galaxies (\citealt{Kauff03, Brinch04}; black contours), the \textsc{agn-host} sample (red circles) and \textsc{inactive} sample (blue contours). {\Rfive The star forming ``main sequence'' fitted by eye to the MPA-JHU catalogue data is shown (solid line) with $\pm 0.3$ dec (dashed lines).}}}
\label{cmdsfms}
\end{figure}

Since this investigation is focussed on whether an AGN can have an impact on the SF of its host galaxy, we must also consider {\changed possible selection effects}. The extent to which SF {\changedbds could obscure AGN} emission was addressed by \cite{Sch2010}. They showed, via analysis of simulated AGN emission added to star-forming galaxies, that BPT-based selection of AGN produces a complete sample at luminosities of $L[OIII] > 10^{40}~\rm{erg~s}^{-1}$. Above this limit we therefore assume we have selected a complete sample of AGN independent of host galaxy SFR.

\subsection{Bayesian SFH Determination}\label{starpy}


\textsc{starpy}\footnote{Publicly available: \url{http://github.com/zooniverse/starpy}} is a \textsc{python} code which allows the user to derive the {\newref quenching} star formation history (SFH) of a single galaxy through a Bayesian Markov Chain Monte Carlo method \citep{Dan}\footnote{\url{http://dan.iel.fm/emcee/}} with the input of the observed $u-r$ and $NUV-u$ colours, a redshift, and the use of the stellar population models of \cite{BC03}. {\changed These models are implemented using solar metallicity (varying this does not substantially affect these results; \citealt{Sme2015}) and a Chabrier IMF \citep{Chab03} {\secondchange but does not model for intrinsic dust (see Section \ref{agnsample})}. The SFH is modelled as an exponential decline of the SFR described by two parameters $[t_q, \tau]$, where $t_q$ is the time at the onset of quenching} $\rm{[Gyr]}$ and $\tau$ is the exponential rate at which quenching occurs $\rm{[Gyr]}$. Under the simplifying assumption that all galaxies formed at $t=0$ $\rm{ Gyr}$ with an initial burst of star formation, the SFH can be described as: 
\begin{equation}\label{sfh}
SFR =
\begin{cases}
i_{sfr}(t_q) & \text{if } t < t_q \\
i_{sfr}(t_q) \times exp{\left( \frac{-(t-t_{q})}{\tau}\right)} & \text{if } t > t_q 
\end{cases}
\end{equation}
where $i_{sfr}$ is an initial constant star formation rate dependent on $t_q$ \citep{Sch2014, Sme2015}.  A smaller $\tau$ value corresponds to a rapid quench, whereas a larger $\tau$ value corresponds to a slower quench. {\newref We note that a galaxy undergoing a slow quench is not necessarily quiescent by the time of observation. Similarly, despite a rapid quenching rate, star formation in a galaxy may still be ongoing at very low rates, rather than being fully quenched. } {\changed This SFH model has previously been shown to appropriately characterise quenching galaxies \citep{Weiner06, Martin07, Noeske07,Sch2014}. {\Rfour We note also that star forming galaxies in this regime are fit by a constant SFR with a $t_q \simeq$ Age$(z)$, (i.e. the age of the Universe at the galaxy's observed redshift) with a very low probability.} }

The probabilistic fitting methods to these star formation histories for an observed galaxy are described in full detail in {\changed Section 3.2 of} \cite{Sme2015}, wherein the \starpy ~~code was used to characterise the SFHs of each galaxy in the \textsc{gz2-galex} sample. {\changed We assume a flat prior on all the model parameters and {\secondchange the difference between the observed and predicted $u-r$ and $NUV-u$ colours are modelled as independent realisations of a double Gaussian likelihood function} (Equation 2 in \citealt{Sme2015}).} {\secondchange We also make the simplifying assumption that the age of each galaxy, $t_\mathrm{age}$ corresponds to the age of the Universe at its observed redshift, $t_\mathrm{obs}$.} 

The output of \starpy  ~ is probabilistic in nature and provides the posterior probability distribution across the two-parameter space for an individual galaxy {\newref the degeneracies for which can be seen in Figure~4 of \citet{Sme2015}}. {\changed To study the SFH across a population of galaxies, these individual posterior probability distributions are {\secondchange stacked in $[t, \tau]$ space {\Rfour and weighted by their probability. This is to minimise the contribution of galaxies poorly fit by this {\Rfive exponentially declining SFH, therefore galaxies in each sample which reside on the main sequence will not contribute to the final population distribution of quenching parameters shown in Figures \ref{time} \& \ref{rate}}. This is no longer inference but is a method to visualise the results across a population of galaxies. An alternative method would be to perform inference on hierarchical Bayesian parent parameters, $\vec{\theta'}$ to describe the population. Such a hierarchical method, however, requires an initial decision on the functional shape of this parent distribution, which introduces non-trivial assumptions. A discussion of this approach and the decision to use an alternative approach can be found in Section \ref{althyper}.}

We obtain separate stacked population distributions for both smooth and disc galaxies by using the GZ2 debiased vote fractions for disc ($p_d$) or smooth ($p_s$) morphologies as weights when stacking, as in \citet{Sme2015}. This ensures that the entirety of the population is used, with galaxies with a higher $p_d$ contributing more to the disc weighted than the smooth weighted population density. This negates the need for a threshold on the GZ2 vote fractions \citep[e.g., $p_d > 0.8$ as used in][]{Sch2014}. {\secondchange These distributions will be referred to as the population densities.}}

We also split both the \textsc{agn-host} and \textsc{inactive} samples into low, medium and high mass ranges (see Table~\ref{massbins}) to investigate any trends in the SFH with mass. {\changed The mass boundaries were chosen to give roughly equal numbers of inactive galaxies in each bin prior to the mass matching to the \textsc{agn-host} sample.}

\section{Results}\label{results}

\begin{figure*}
\includegraphics[width=0.82\textwidth]{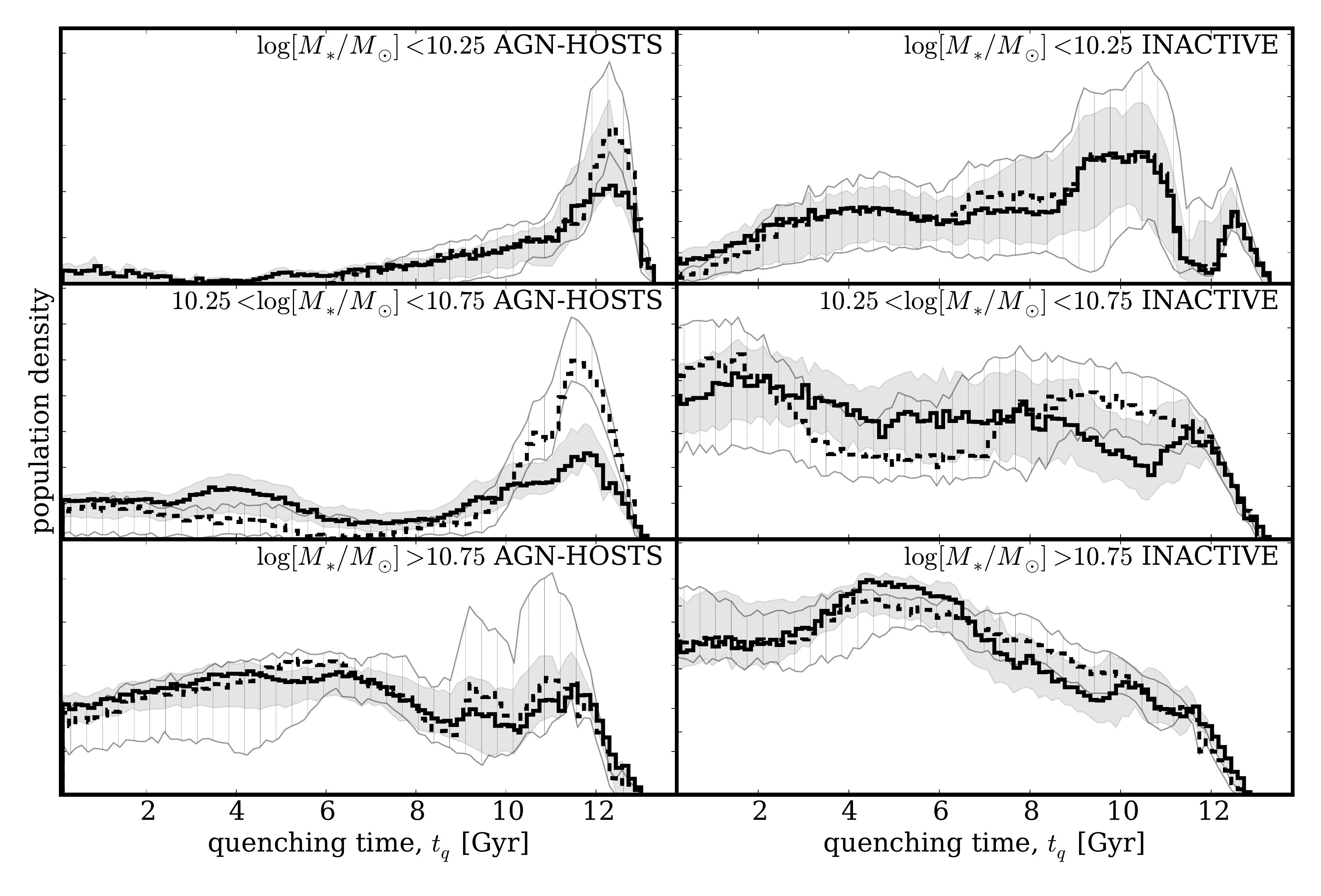}
\caption{{\secondchange Population density} distributions for the quenching time ($t_q$) parameter, {\changedbds normalised so that the areas under the curves are equal}. \textsc{agn-host} (left) and \textsc{inactive} (right) galaxies are split into low (top), medium (middle) and high (bottom) mass for smooth (dashed) and disc (solid) galaxies. {\newref Uncertainties from bootstrapping are shown by the shaded regions for the smooth (grey striped) and disc (grey solid) population densities.} A low (high) value of $t_q$ corresponds to the early (recent) Universe.}
\label{time}
\end{figure*}

\begin{figure*}
\includegraphics[width=0.82\textwidth]{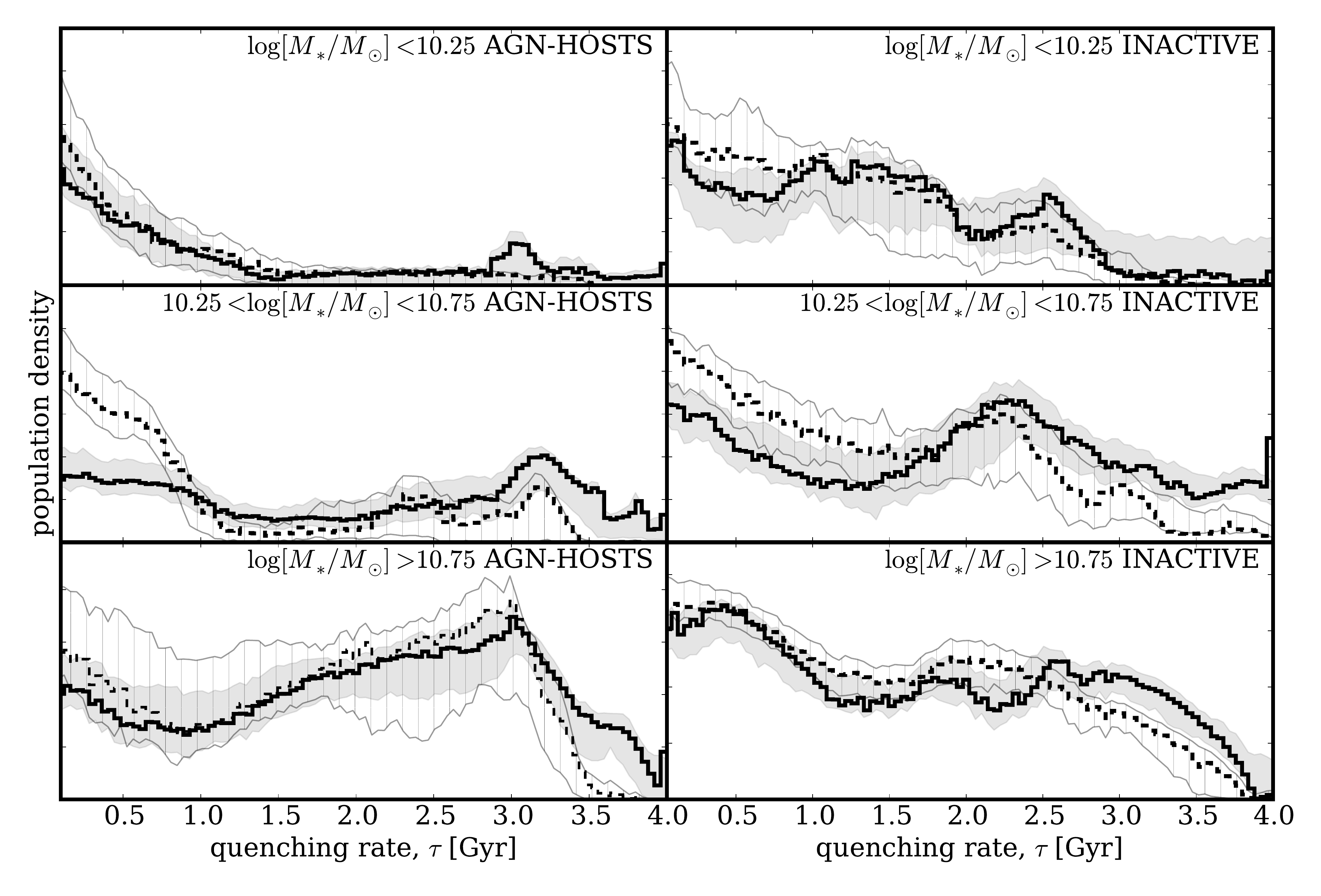}
\caption{{\secondchange Population density} distributions for the quenching rate ($\tau$), normalised so that the areas under the curves are equal. \textsc{agn-host} (left) host and \textsc{inactive} (right) galaxies are split into low (top), medium (middle) and high (bottom) mass for smooth (dashed) and disc (solid) galaxies. {\newref Uncertainties from bootstrapping are shown by the shaded regions for the smooth (grey striped) and disc (grey solid) population densities.} A small (large) value of $\tau$ corresponds to a rapid (slow) quench.}
\label{rate}
\end{figure*}

\begin{table*}
\centering
\caption{Table showing the number of galaxies in each of the three mass bins for both the \textsc{agn-hosts} and \textsc{inactive} galaxy samples and the percentage of the {\secondchange distribution} across each morphologically weighted population found in the rapid, intermediate and slow quenching regimes.}
\label{massbins}
\begin{tabular}{c|c|c|c|c|c|c}
\hline
\textsc{sample}                     & \textsc{mass bin}                                        & \textsc{weighting}                  & $\tau < 1 ~\rm{[Gyr]}$                             & $1 < \tau ~\rm{[Gyr]} < 2 $          & $\tau > 2 ~\rm{[Gyr]}$                               & \textsc{number}                                        \\ \hline \hline
\multirow{6}{*}{AGN-HOSTS} & \multirow{2}{*}{$\log [M_*/M_{\odot}] < 10.25 $}                       & $p_d$     & $60\pm_{5}^{23}$                    & $13\pm_{9}^{9}$                    & $28\pm_{19}^{6}$       & \multirow{2}{*}{$165 (13.3\%)$}                      \\
                           &                                                 & $p_s$     & $69\pm_{6}^{14}$                    & $17\pm_{14}^{6}$                   & $14\pm_{7}^{3}$        &                                                      \\ \cline{2-7} 
                           & \multirow{2}{*}{$10.25 < \log [M_*/M_{\odot}] < 10.75$}                    & $p_d$     & $33\pm_{5}^{3}$                     & $15\pm_{4}^{4}$                    & $51\pm_{7}^{4}$        & \multirow{2}{*}{$630 (50.6\%)$}                      \\
                           &                                                 & $p_s$     & $69\pm_{5}^{4}$                     & $7\pm_{4}^{4}$                     & $26\pm_{9}^{5}$        &                                                      \\ \cline{2-7} 
                           & \multirow{2}{*}{$\log [M_*/M_{\odot}] > 10.75$}                      & $p_d$     & $20\pm_{4}^{5}$ & $25\pm_{5}^{7}$                    & $56\pm_{12}^{8}$       & \multirow{2}{*}{$449 (36.1\%)$}                      \\
                           &                                                 & $p_s$     & $24\pm_{3}^{4}$                     & $26\pm_{6}^{5}$                    & $50\pm_{7}^{7}$        &                                                      \\ \hline \hline
\multirow{6}{*}{INACTIVE}  & \multirow{2}{*}{$\log [M_*/M_{\odot}] < 10.25 $}                       & $p_d$     & $37\pm_{14}^{8}$                    & $39\pm_{6}^{8}$                    & $24\pm_{6}^{8}$        & \multirow{2}{*}{$807 (13.2\%)$}                      \\
                           &                                                 & $p_s$     & $47\pm_{11}^{5}$                    & $36\pm_{5}^{9}$                    & $17\pm_{5}^{4}$        &                                                      \\ \cline{2-7} 
                           & \multirow{2}{*}{$10.25 < \log [M_*/M_{\odot}] < 10.75$}                    & $p_d$     &          $30\pm_{3}^{4}$                          &       $18\pm_{3}^{2}$                            &    $51\pm_{4}^{4}$                   & \multirow{2}{*}{$3094 (50.7\%)$}                     \\
                           &                                                 & $p_s$     & $42\pm_{2}^{2}$            & $29\pm_{3}^{3}$   & $30\pm_{4}^{3}$ &                                                      \\ \cline{2-7} 
                           & {\multirow{2}{*}{$\log [M_*/M_{\odot}] > 10.75$}} & $p_d$     & $36\pm_{3}^{3}$            & $24\pm_{4}^{3}$         & $41\pm_{3}^{4}$ & \multicolumn{1}{l}{\multirow{2}{*}{$2206 (36.1\%)$}} \\
                           & \multicolumn{1}{l|}{}                           & $p_s$      & $38\pm_{2}^{2}$              & $28\pm_{4}^{3}$            & $34\pm_{3}^{3}$ & \multicolumn{1}{l}{}                                 \\ \hline                       
\end{tabular}
\end{table*}

Figures~\ref{time} and \ref{rate} show the stacked {\secondchange population density} distributions for the quenching time, $t_q$ and exponential quenching rate, $\tau$, respectively. In each figure the {\secondchange population density}, along with shaded regions to show the uncertainties, for a given parameter is shown for smooth and {\changed disc galaxy populations} across three mass bins for the \textsc{agn-host} and \textsc{inactive} samples. {\Rfive No cut on the star formation rate is made to the galaxies which contribute to Figures \ref{time} \& \ref{rate}, but those galaxies poorly fit by an exponentially declining SFH are down-weighted so that they do not contribute to the results presented here.} {\changed In Table~\ref{massbins} the percentage of the {\secondchange population density} in each quenching regime} for rapid ($\tau < 1$ Gyr), intermediate ($1 < \tau ~\rm{[Gyr]} < 2$) and slow ($\tau > 2$ Gyr) quenching timescales, are shown. {\newref Uncertainties on the population densities (shown by the shaded regions) are determined from the maximum and minimum values spanned by $N = 1000$ bootstrap iterations, each sampling $90\%$ of the galaxy population. $1\sigma$ uncertainties are quoted for the percentages in Table~\ref{massbins}, calculated from the bootstrapped distributions.}

 {\Rfive These population densities should be interpreted as the spread of quenching times and rates occurring in galaxies which have either undergone or are undergoing quenching within a population}. {\changed Figures~\ref{time} and \ref{rate} show} a distinct difference between the {\secondchange population density} of \textsc{agn-host} and \textsc{inactive} quenching parameters.

At all masses, the {\secondchange population density} for {\Rfive galaxies within} the \textsc{agn-host} population across the quenching time $t_q$ parameter (left panels of Figure~\ref{time}) is different from that of the inactive galaxies (right panels of Figure~\ref{time}). Recent quenching ($t > 11$ Gyr) is the dominant history for {\Rfive quenched and quenching} low and medium mass \textsc{agn-host} galaxies, particularly for the smooth galaxies hosting an AGN. However, this effect is less dominant in higher mass galaxies where quenching at earlier times also has {\secondchange high density}.

The {\secondchange population densities} for the quenching rate, $\tau$, in Figure~\ref{rate} and Table~\ref{massbins} show the dominance of rapid quenching ($\tau < 1$ Gyr) {\Rfive within} the \textsc{agn-host} population, particularly for smooth galaxies. With increasing mass the dominant quenching rate becomes slow ($\tau > 2$ Gyr) especially for disc galaxies hosting an AGN. Similar trends in the {\secondchange density} are observed {\Rfive within the} \textsc{inactive} population but the overall distribution is very different.

The {\secondchange distributions} for the \textsc{agn-host} galaxies therefore show evidence for the dominance of rapid, recent quenching {\Rfive within} this population. This result implies the importance of AGN feedback for the evolution of these galaxies.

\section{Discussion}\label{dis}

The differences between the {\secondchange population density distributions} of the \textsc{agn-host} and \textsc{inactive} populations reveal that an AGN can have a significant effect on the SFH of its host galaxy. Both recent, rapid quenching and early, slow quenching are observed in the {\secondchange population density} {\Rfive within} the \textsc{agn-host} population.

{\changed There are minimal differences between the smooth and disc weighted distributions of the quenching parameters {\Rfive within} the \textsc{agn-host} population. This is agreement with the conclusions of \citet*{Kauff03b} who found that the structural properties of AGN hosts depend very little on AGN power. }

The difference between the \textsc{agn-host} and \textsc{inactive} {\secondchange population distributions} in Figure~\ref{rate} for the rate of quenching, $\tau$, tells a story of gas reservoirs. The {\secondchange density distribution} for higher mass \textsc{agn-host} galaxies is dominated by slow, early quenching implying another mechanism is responsible for the cessation of star formation {\Rfive within} {\Rfour a proportion of} these high mass galaxies prior to the triggering of the current AGN.  This preference for slow evolution timescales follows from the ideas of previously isolated discs evolving slowly by the Kennicutt-Schmidt \citep{Schmidt59, Kennicutt97} law which can then undergo an interaction or merger to reinvigorate star formation, feed the central black hole and trigger an AGN \citep{Varela04, Em15}. These galaxies would need a large enough gas reservoir to fuel both SF throughout their lifetimes and the recent AGN. These high mass galaxies also play host to the most luminous AGN (mean $\log (L[OIII] ~[\rm{erg}~s^{-1}]) \sim 41.6$) and so this SFH challenges the usual explanation for the co-evolution of luminous black holes and their host galaxies driven by merger growth.

Quenching at early times is also observed {\Rfive within} a subsample of the \textsc{inactive} population, where the {\secondchange density} for the quenching time is roughly constant until recent times where the distribution drops off. {\changed This drop-off occurs at earlier times with increasing mass with a significant lack of quenching occurring at early times for low mass \textsc{inactive} galaxies} (right panels Figure~\ref{time}). This is evidence of downsizing {\Rfive within} the \textsc{inactive} galaxy population whereby stars in massive galaxies form first and quench early \citep{Cowie96, Thomas10}. 

{\Rfour Some of the} most massive \textsc{agn-host} galaxies also show a preference for earlier quenching (bottom left panel Figure~\ref{time}) occurring at slow rates; we speculate that this is also due to the effects of downsizing rather than being caused by the current AGN. This earlier evolution would first form a slowly `dying' or `dead' galaxy typical of massive elliptical galaxies which can then have a recent infall of gas either through a minor merger, galaxy interaction or environmental change, triggering further star formation and feeding the central black hole, triggering an AGN \citep{Kav14}. In turn this AGN can then quench the recent boost in star formation. This track is similar to the evolution history proposed for blue ellipticals \citep{Kav13, McIntosh14, Haines15}. This SFH would then give rise to the distribution seen {\Rfive within} the high mass \textsc{agn-host} population for both time and rate parameters.

These recently triggered AGN in both massive disc and smooth galaxies do not have have the ability to impact the SF across the entirety of a high mass galaxy in a deep gravitational potential \citep{Ish12, Zinn13}. This leads to the lower peak for recent, rapid quenching {\Rfive within} the high mass \textsc{agn-host} population for both morphologies. 

Conversely, rapid quenching, possibly caused by the AGN itself through negative feedback, is the most dominant history {\Rfive within} the low mass \textsc{agn-host} population with lower gravitational potentials from which gas may be more readily expelled or heated \citep{Torbra09}. 

\cite{Torbra09} model the effects of jet-induced AGN feedback on a typical early type (i.e. smooth) galaxy {\changed and observe a drastic suppression of star formation on a timescale of $\sim 3 ~\rm{Myr}$. Comparing their synthetic colours with observed colours of SDSS elliptical galaxies, they} find the time between the current galaxy age, $t_\mathrm{gal}$ and the time that the feedback began, $t_\mathrm{AGN}$, peaks at $t_\mathrm{gal} - t_\mathrm{AGN} \sim 0.85 ~\rm{Gyr}$. This agrees with the location of the peak in Figure~\ref{time} for low mass galaxies {\Rfive which have undergone quenching}, where the difference between the peak of the {\secondchange distribution} and the average age of the population ({\secondchange galaxy age is calculated as the age of the Universe at the observed redshift}, by assuming all galaxies form at $t=0$) is $\sim0.83 ~\rm{Gyr}$. This implies that this SFH dominated by recent quenching is caused directly by negative AGN feedback.

{\changed However, there still remains the possibility that the AGN is merely a consequence of an alternative quenching mechanism. This idea is supported by simulations showing that the exhaustion of gas by a merger fuelled starburst could cause such a rapid quench in star formation and in turn also trigger an AGN \citep{Croton06, Wild09, Snyder11, Hayward14}. \citet{Yesuf14} also showed that AGN are more commonly hosted by post starburst galaxies, with the peak AGN activity appearing $\geq 200 \pm 100 ~\rm{Myr}$ after the starburst. Such a SFH is not accounted for in the models presented here, however this scenario is still consistent with the results presented in this paper; that AGN which are \emph{currently} active have been detected in host galaxies $\sim 1~\rm{Gyr}$ after the onset of quenching.}

This rapid quenching is particularly dominant for low-to-medium mass smooth galaxies. \cite{Sme2015} suggest that incredibly rapid quenching rates could be attributed to mergers of galaxies in conjunction with AGN feedback, which are thought to be responsible for creating the most massive smooth galaxies \citep{Con03, SdMH05, Hopkins08}. This dominance of rapid quenching across the smooth \textsc{agn-host} population supports the idea that a merger, having caused a morphological transformation to a smooth galaxy, can also trigger an AGN, causing feedback and cessation of star formation (\citealt{Sanders88}).

{\Rfive Within} the medium mass \textsc{agn-host} population we see a bimodal distribution between these two quenching histories, highlighting the strength of this method which is capable of detecting such variation in the SFHs {\Rfive within} a population of galaxies. 

{\Rfive Indeed not all galaxies in the \textsc{agn-host} and \textsc{inactive} samples are quenching, as seen in Figure \ref{cmdsfms}, with a significant proportion of both the \textsc{agn-host} and \textsc{inactive} samples lying on the star forming sequence. A galaxy can therefore still maintain star formation whilst hosting an AGN. The results presented in Section \ref{results} only reflect the trends for galaxies that have undergone or are currently undergoing quenching within a population and can therefore be accurately fit by an exponentially declining SFH. This prevalence of star forming AGN host galaxies, combined with the results above allows us to consider that either: (i)  the AGN are the cause of the rapid quenching observed but only in gas-poor host galaxies where they can have a large impact, (ii) the AGN are a consequence of another quenching mechanism but can also be triggered by other means which do not cause quenching, or (iii) the SFR of a galaxy can recover post-quench and return to the star forming sequence after a few Gyr (see recent simulations by \citealt{Pontzen16}). Further investigation will therefore be required to determine the nature of this quenching.}
\\
\\
We have used morphological classifications from the Galaxy Zoo 2 project to determine the morphology-dependent SFHs of a population of $1,244$ Type 2 Seyfert AGN host galaxies, in comparison to an inactive galaxy population, via a {\secondchange partially} Bayesian analysis of an exponentially declining SFH model. We determined the {\secondchange population distribution} for the quenching onset time, $t_q$, and exponential quenching rate, $\tau$, and find clear differences in the {\secondchange distributions}, between inactive and AGN host galaxy populations. We have demonstrated a clear dependence on a galaxy currently hosting an AGN and its {\Rfive SFH for those galaxies which have undergone or are undergoing quenching}. There is strong evidence for downsizing in massive inactive galaxies, which appears as a secondary effect in AGN host galaxies. The dominant mechanism for {\Rfive quenched and quenching} galaxies currently hosting an AGN is for rapid quenching which has occurred very recently. This result demonstrates the importance of AGN feedback {\Rfive within} the host galaxy population, in driving the evolution of galaxies across the colour-magnitude diagram.

\section*{Acknowledgements}

The authors would like to thank S. Kaviraj and P. Marshall for helpful discussions. We also thank {\newref both of the} anonymous referees for helpful comments that greatly improved the clarity of the paper. 

RJS acknowledges funding from the STFC Grant Code ST/K502236/1. BDS gratefully acknowledges support from Balliol College, Oxford. KS gratefully acknowledges support from Swiss National Science Foundation Grant PP00P2 138979/1. SJK acknowledges funding from the STFC Grant Code ST/MJ0371X/1. KLM acknowledges funding from The Leverhulme Trust as a 2010 Early Career Fellow. KWW acknowledges funding from NSF grant AST-1413610. OIW acknowledges a Super Science Fellowship from the Australian Research Council. Support for this work was provided by the National Aeronautics and Space Administration through Einstein Postdoctoral Fellowship Award Number PF5-160143 issued by the Chandra X-ray Observatory Center, which is operated by the Smithsonian Astrophysical Observatory for and on behalf of the National Aeronautics Space Administration under contract NAS8-03060. The development of Galaxy~Zoo was supported by the Alfred P. Sloan Foundation and The Leverhulme Trust. Based on observations made with the NASA GALEX\footnote{\url{http://galex.stsci.edu/GR6/}} and the SDSS\footnote{\url{https://www.sdss3.org/collaboration/boiler-plate.php}}.


{}

\appendix

\section{Mass matched INACTIVE sample}

\begin{figure*}
\includegraphics[width=0.8\textwidth]{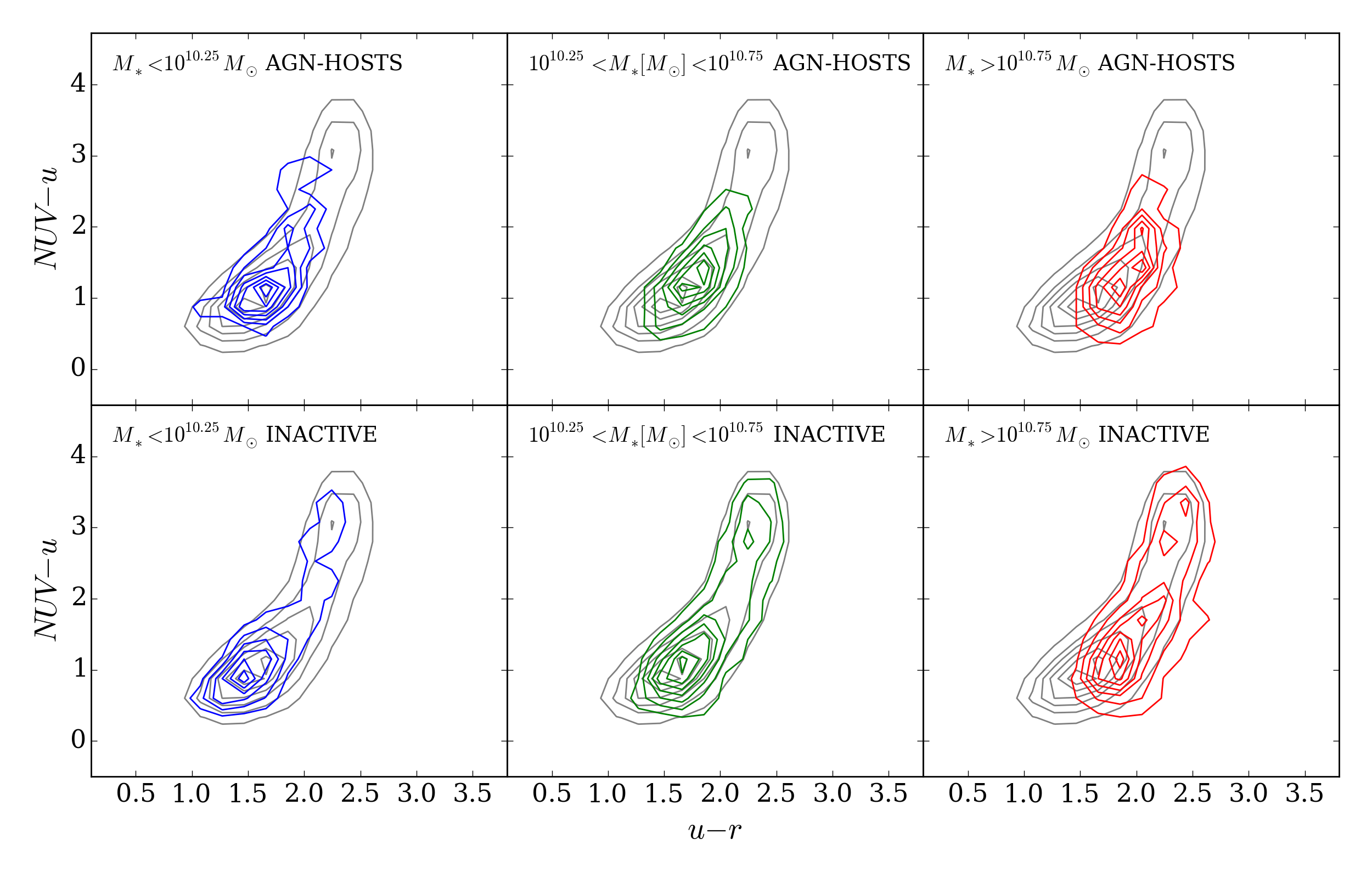}
\caption{Optical-NUV colour-colour contour diagrams for the \textsc{agn-host} (top) and \textsc{inactive} galaxy samples split into low (blue), medium (green) and high (red) stellar mass samples. Underlaying each diagram are the contours of the \textsc{gz2-galex} sample (grey).}
\label{colcol}
\end{figure*}

Each galaxy in the \textsc{agn-host} sample has been matched to at least one and up to five inactive galaxies. These were matched to within $\pm5\%$ of the stellar mass and $\pm 0.1$ of each of the disc and smooth GZ2 vote fractions, $p_d$ and $p_s$.

Both the \textsc{agn-host} and \textsc{inactive} galaxy samples are shown on an optical-NUV colour colour diagram in Figure~\ref{colcol}. The \textsc{inactive} sample across all mass bins can be seen to encompass the entirety of the colour magnitude diagram, unlike the \textsc{agn-host} sample which reside at increasingly green colours with increasing mass.

\section{Luminosity dependence}

\begin{figure}
\includegraphics[width=0.35\textwidth]{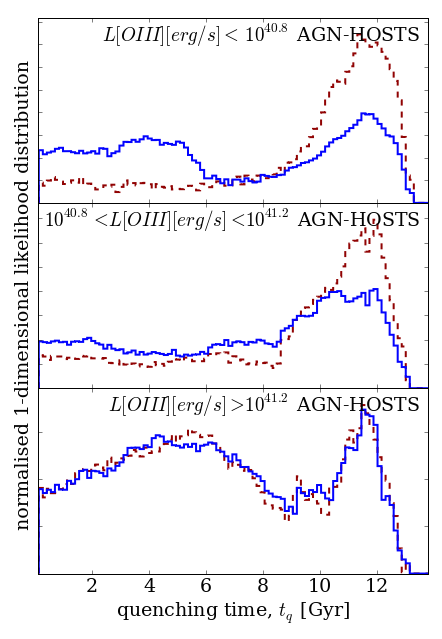}
\caption{{\secondchange Population density} distributions for the quenching time, $t_q$, {\changedbds normalised so that the areas under the curves are equal}. \textsc{agn-host} galaxies are split into low (top), medium (middle) and high (bottom)  $L[OIII]$ for smooth (red dashed) and disc (blue solid) galaxies. A low (high) value of $t_q$ corresponds to the early (recent) Universe.}
\label{loiiitime}
\end{figure}

\begin{figure}
\includegraphics[width=0.35
\textwidth]{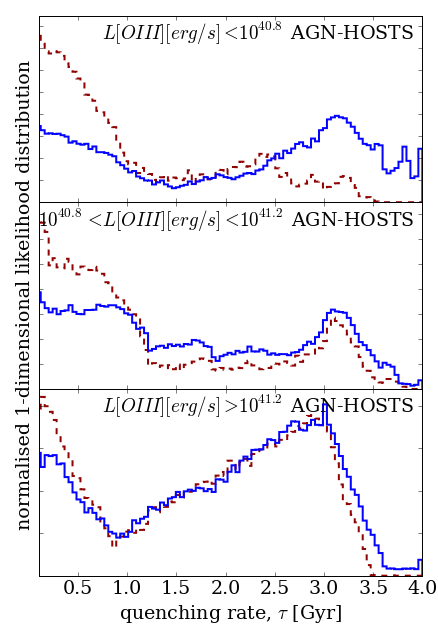}
\caption{{\secondchange Population density} distributions for the quenching rate, $\tau$ normalised so that the areas under the curves are equal. \textsc{agn-host} galaxies are split into low (top), medium (middle) and high (bottom) $L[OIII]$ for smooth (red dashed) and disc (blue solid) galaxies. A small (large) value of $\tau$ corresponds to a rapid (slow) quench.}
\label{loiiirate}
\end{figure}

An investigation into the dependence of the quenching in the \textsc{agn-host} sample with $L[OIII]$ was also conducted, with the summed weighted {\secondchange population density} distributions for the quenching time and rate parameters shown in Figures~\ref{loiiitime} \& \ref{loiiirate}. The \textsc{agn-host} sample was split into low, medium and high luminosity as with the stellar mass. Since the  $L[OIII]$ of the black hole is correlated to the accretion rate \citep{Kauff03b}, which is dependent on the black hole mass, which is in turn correlated to the mass of the host galaxy \citep{Mag98}, the stellar mass was used in the main investigation to allow a direct comparison to the control \textsc{inactive} sample.

\section{Alternative Hierarchical Bayesian approach}\label{althyper}

{\secondchange The approach used in the paper to present the main results shown in Figures~\ref{time} \& \ref{rate}, relied upon a visualisation of the SFHs across each population, with no inference involved beyond the use of \textsc{starpy} to derive the individual galaxy SFHs. The preferred approach to this problem would be to use a hierarchical Bayesian method to determine the `hyper-parameters' that describe the distribution of the parent population $\theta' = [t_q', \tau']$ that each individual galaxy's SFH is drawn from. 

We want the posterior PDF for $\vec{\theta}'$ to describe a galaxy population:
\begin{equation}\label{hyper}
P(\vec{\theta}'|\vec{d}) = \frac{P(\vec{d}|\vec{\theta}')P(\vec{\theta}')}{P(\vec{d})}, 
\end{equation}
where $\vec{d}$ represents all of the optical and NUV colour data in a population $\{\vec{d}_k\}$. For one galaxy, $k$, the marginalised likelihood is:
\begin{equation}\label{one}
P(d_k|\vec{\theta}') = \iint \! P(d_k|t_k, \tau_k) P(t_k, \tau_k|\vec{\theta}') \ \mathrm{d}t_k ~ \mathrm{d}\tau_k
\end{equation}
and for all galaxies, $N$, therefore: 
\begin{equation}
P(\vec{d}|\vec{\theta}') = \prod_k^N P(d_k|\vec{\theta}').
\end{equation}

Using \textsc{starpy}~ for individual galaxies we sample from the `interim' posterior $P(t_k, \tau_k|d_k)$ which we can relate to $P(d_k|t_k, \tau_k)$  so that:

\begin{equation}\label{marg}
P(d_k|\vec{\theta}') = \iint  \! P(t_k, \tau_k|d_k) . P(d_k) . \frac{P(t_k, \tau_k|\vec{\theta}')}{P(t_k, \tau_k)} \ \mathrm{d}t_k ~ \mathrm{d}\tau_k.
\end{equation}
In order to calculate this we draw $N_s$ random samples, $r$, from the interim posterior, $P(t_k, \tau_k|d_k)$ so that Equation \ref{marg} can be expressed as a sum over a number of random samples, $N_s$ (as with the calculation of an expected mean):
\begin{equation}\label{imp}
P(d_k|\vec{\theta}') = \frac{P(d_k)}{N_s} \sum_r^{N_s} \frac{P(t_{k,r}, \tau_{k,r}|\vec{\theta}')}{P(t_k, \tau_k)},
\end{equation}
for the $r^{th}$ sample of $N_s$ total samples taken from one galaxy's, $k$,  interim posterior PDF. This fraction is known as the `importance weight', $w_r$, in importance sampling. 

However, we also have two morphological vote fractions that we can weight by to determine separate hyper-parameters, $\vec{\theta}' = [\vec{\theta}'_d, \vec{\theta}'_s]$, for both disc, $d$, and smooth, $s$, galaxies. Therefore:

\begin{equation}\label{morphimp}
w_r = \frac{P(t_{k,r}, \tau_{k,r}|\vec{\theta}')}{P(t_k, \tau_k)} =  \frac{p_{d,k} P(t_{k,r}, \tau_{k,r}|\vec{\theta}'_d) + p_{s,k} P(t_{k,r}, \tau_{k,r}|\vec{\theta}'_s)}{P(t_k, \tau_k)}
\end{equation} 

If we substitute equation \ref{imp} into equation \ref{hyper} we find that the $P(d_k)$ terms cancel and we are left with:
\begin{equation}
P(\vec{\theta}'|\vec{d}) = P(\vec{\theta}')~\prod_k^N \frac{1}{N_{s,k}} \sum_r^{N_s} w_r ,
\end{equation}
where $P(\vec{\theta}')$ is the assumed prior on the hyper-parameters, which is assumed to be uniform.

This approach is heavily dependent on what shape is assumed for the hyper-distribution; a decision which is not trivial. It is often common for this function to take the form of a multi-component Gaussian mixture model \citep{MacKay, Lahav00}. For example a two component Gaussian mixture model in $[t, \tau]$ space is described by eight hyper-parameters for a single morphology, $\vec{\theta}' = [\mu_{t,1}, \sigma_{t,1}, \mu_{\tau,1}, \sigma_{\tau,1}, \mu_{t,2}, \sigma_{t,2}, \mu_{\tau,2}, \sigma_{\tau,2}]$. Here we also assume no covariance between hyper-parameters for simplicity. 

We used this assumption of a two component Gaussian mixture model, to infer the population parameters for both the \textsc{agn-host} and \textsc{inactive} populations and the results are shown in Figure~\ref{method3}. These results were produced by drawing $N_s = 100$ random samples from each galaxy, $k$, in each mass bin. We plot the distributions for a given morphology by taking the median value of the posterior distribution for each of the 8 parameters describing the two component Gaussian mixture. We can see in Figure~\ref{method3} that this hierarchical method produces similar distributions for the \textsc{agn-host} and \textsc{inactive} samples. This finding is not expected given the differences between the two samples in colour-colour space seen in Figure~\ref{colcol}. 

\begin{figure}
\includegraphics[width=0.48\textwidth]{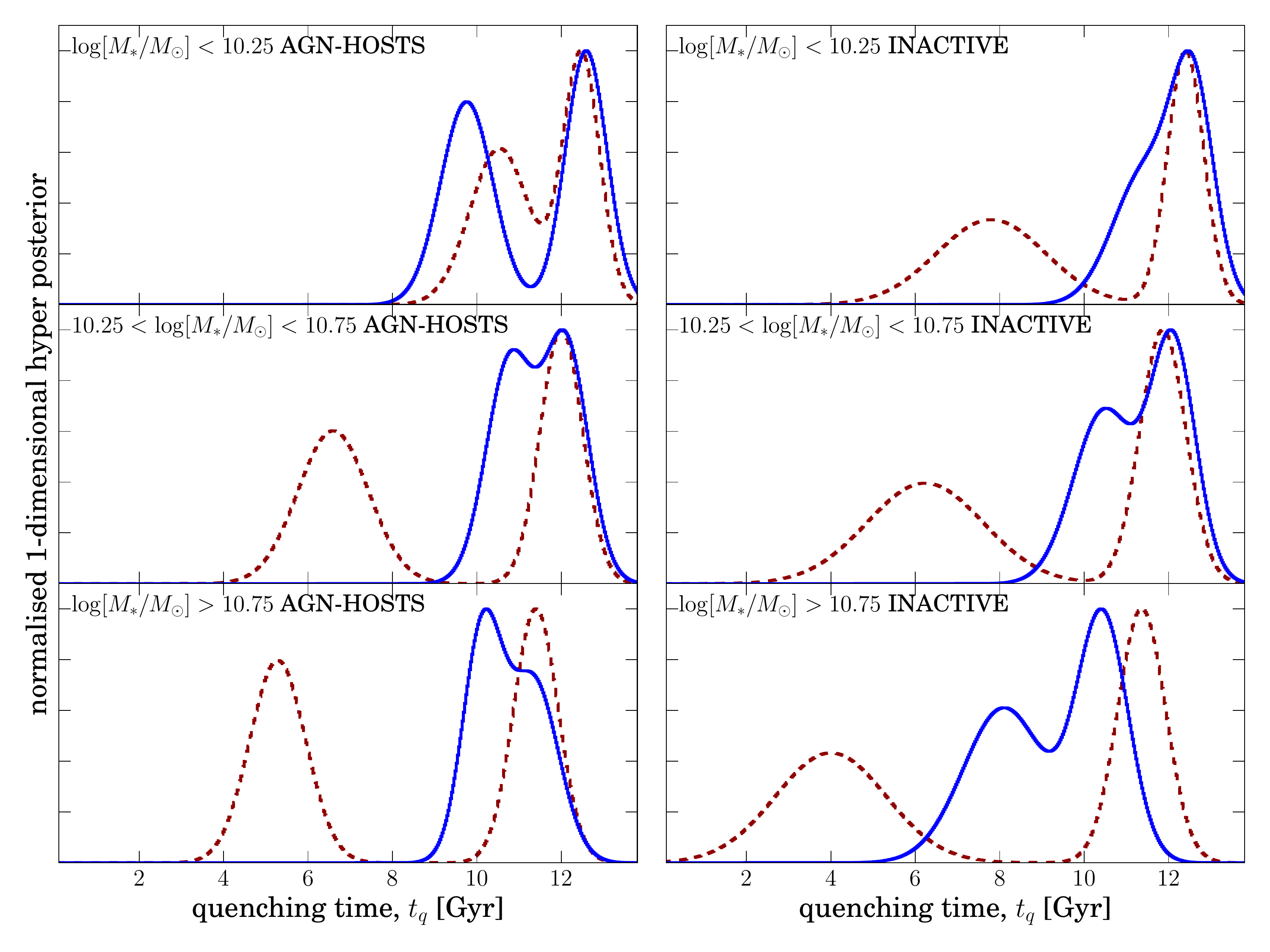}
\includegraphics[width=0.48\textwidth]{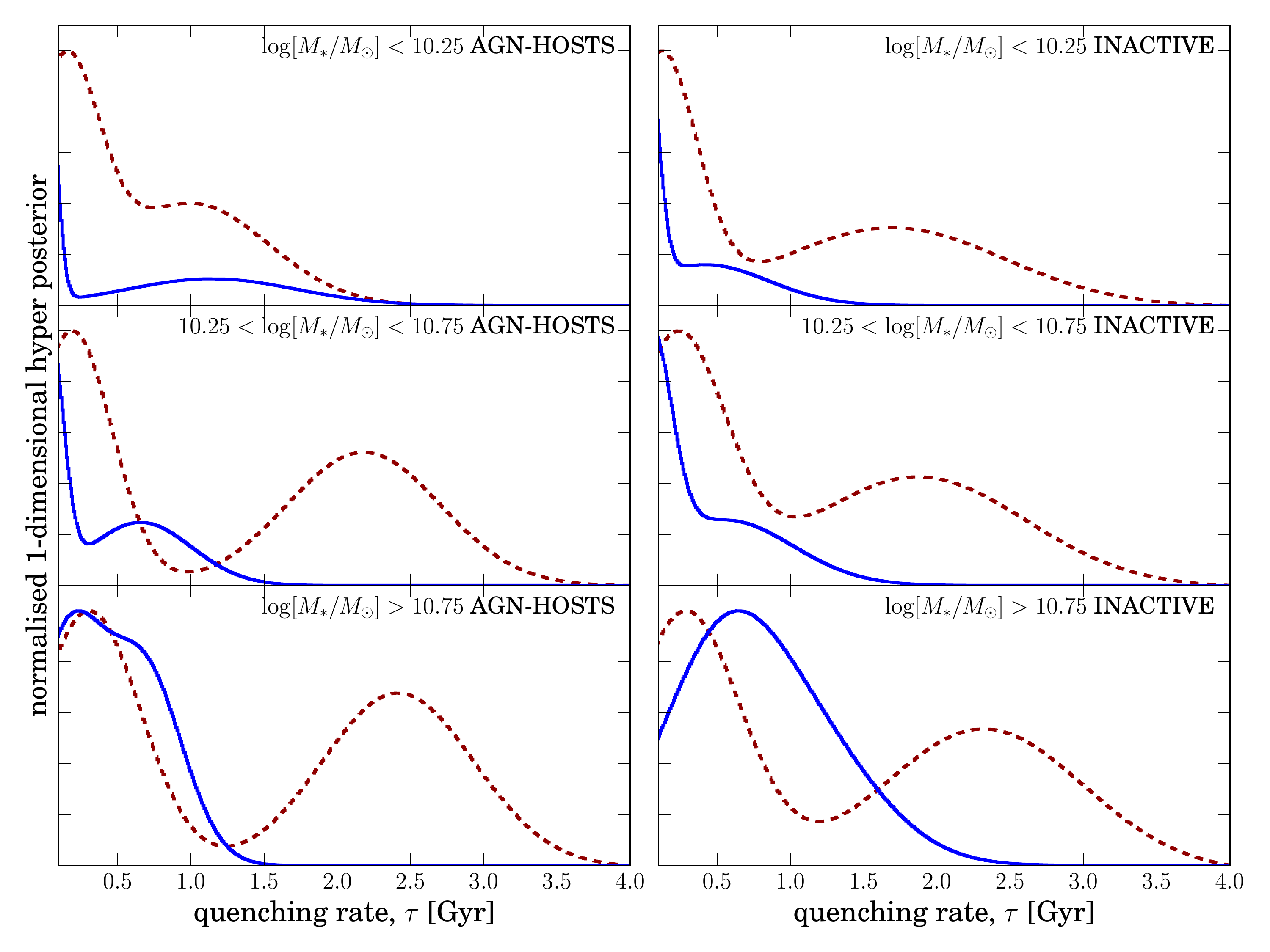}
\caption[8pt]{Hierarchical-posterior PDF of the quenching time ($t_q'$, top) and rate ($\tau'$, bottom) population parameters, normalised so that the areas under the curves are equal. \textsc{agn-host} (left) and \textsc{inactive} (right) galaxies are split into low (top), medium (middle) and high (bottom) mass, weighted for smooth (red dashed) and disc (blue solid) galaxies. A low (high) value of $t_q'$ corresponds to the early (recent) Universe. A small (large) value of $\tau'$ corresponds to a rapid (slow) quench.}
\label{method3}
\end{figure}

In order to test whether this assumption of a multi-component Gaussian mixture model is appropriate, we sampled the inferred hierarchical distributions to produce replica datasets in optical-NUV colour space. These are shown here in Figure~\ref{replica}  in comparison to the observed colour-colour distributions of the \textsc{agn-host} and \textsc{inactive} samples. For all masses and morphologies the replicated $u-r$ and $NUV-u$ colours do not accurately match the observed data. 

\begin{figure}
\begin{centering}
\includegraphics[width=0.42\textwidth]{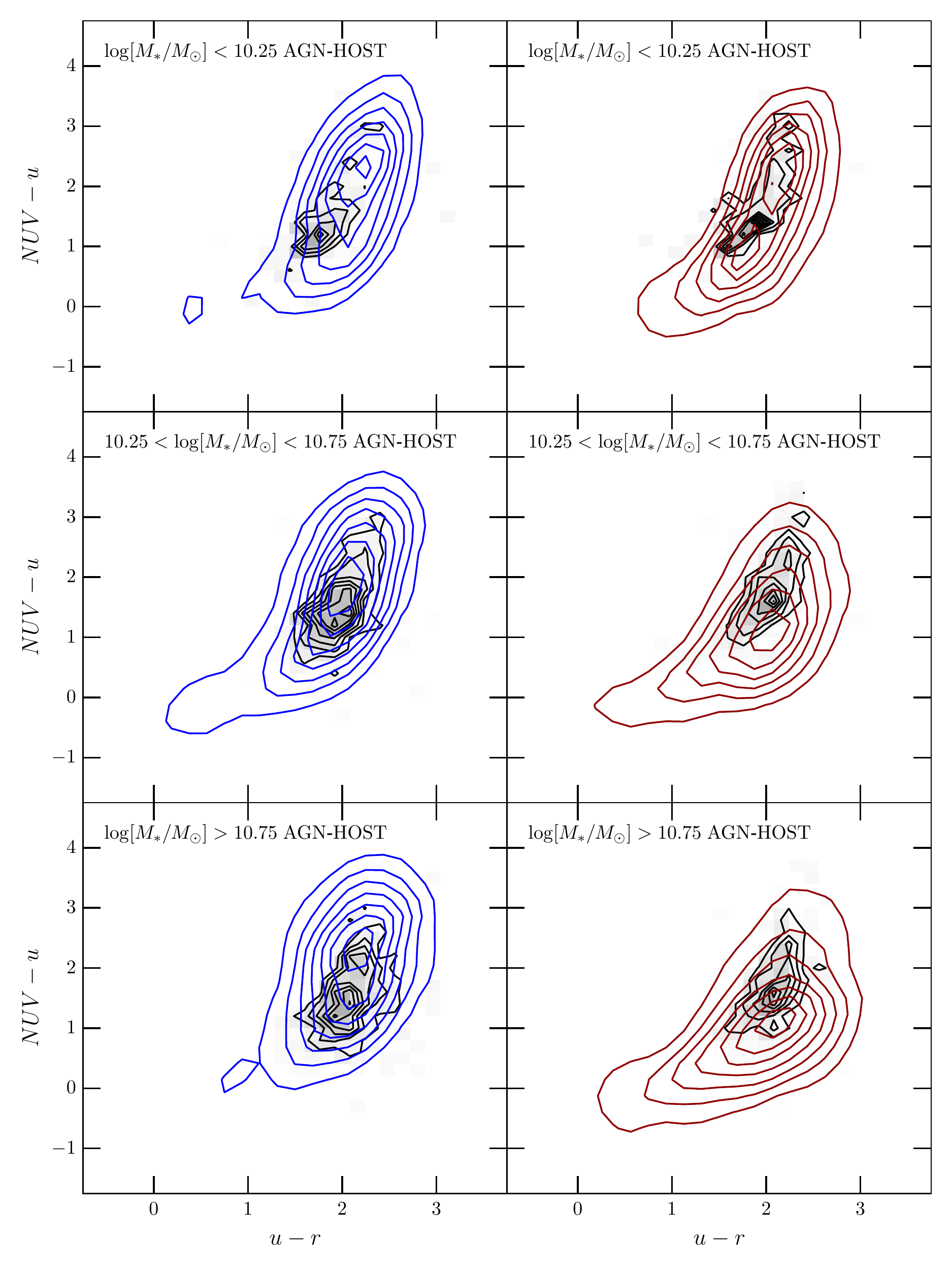}
\includegraphics[width=0.42\textwidth]{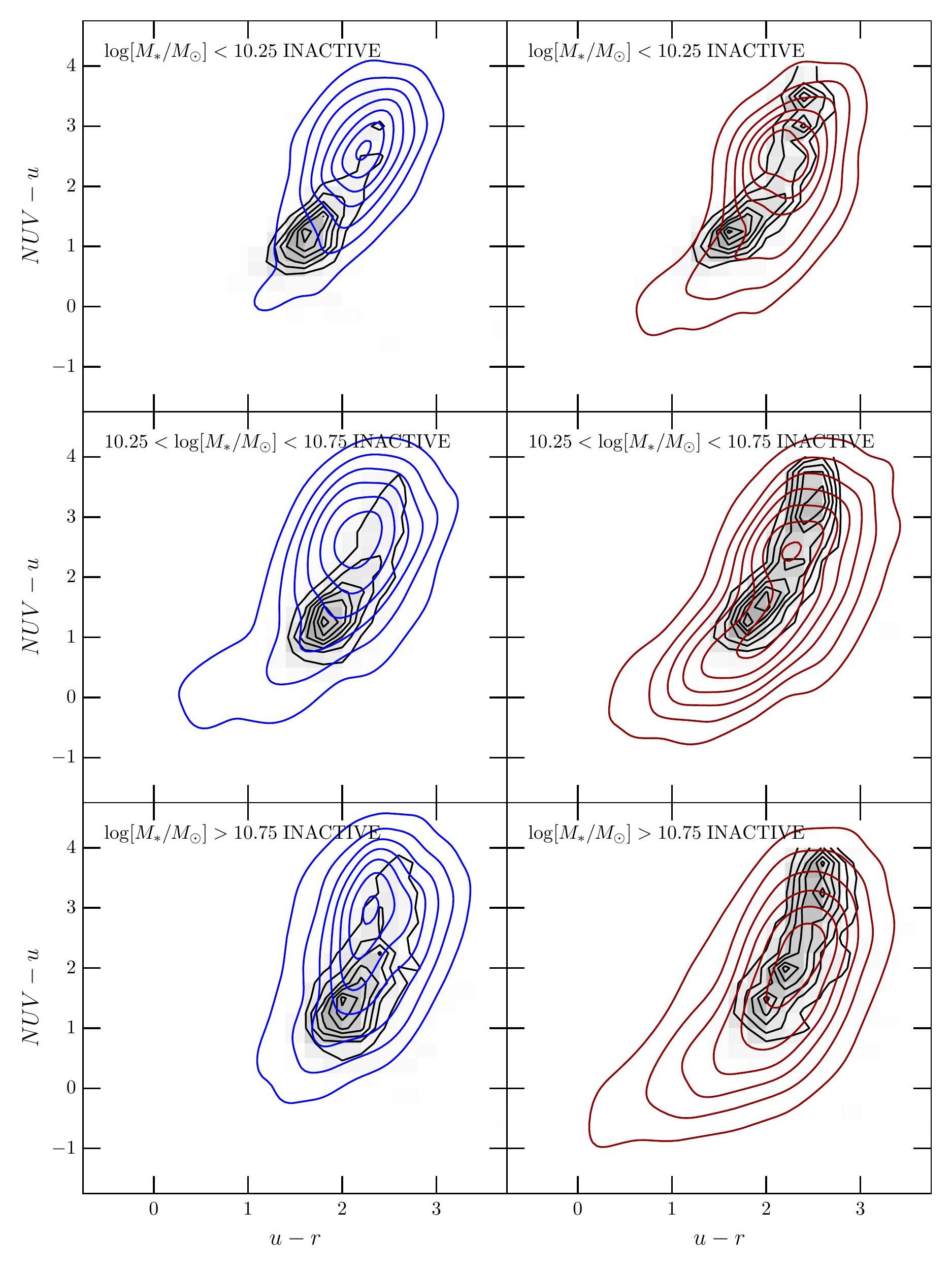}
\caption[8pt]{Optical-NUV colour-colour diagrams for the \textsc{agn-host} (top) and \textsc{inactive} (bottom) galaxies shown by the black contours, split into low mass (top), medium mass (middle) and high mass (bottom) galaxies weighted by $p_d$ (left) and $p_s$ (right). Kernel smoothing has been applied to the overlaid replica datasets, which are created by sampling from the \textbf{inferred 2 component Gaussian mixture model hierarchical parent distributions}, shown here in Figure~\ref{method3}. Gaussian random noise is also added to the inferred colours, with a mean and standard deviation of the errors on the observed colours of the respective sample. Contours are shown for samples taken from the disc (blue) and smooth weighted (red) inferred hierarchical distributions.
}
\label{replica}
\end{centering}
\end{figure}

\begin{figure}
\begin{centering}
\includegraphics[width=0.42\textwidth]{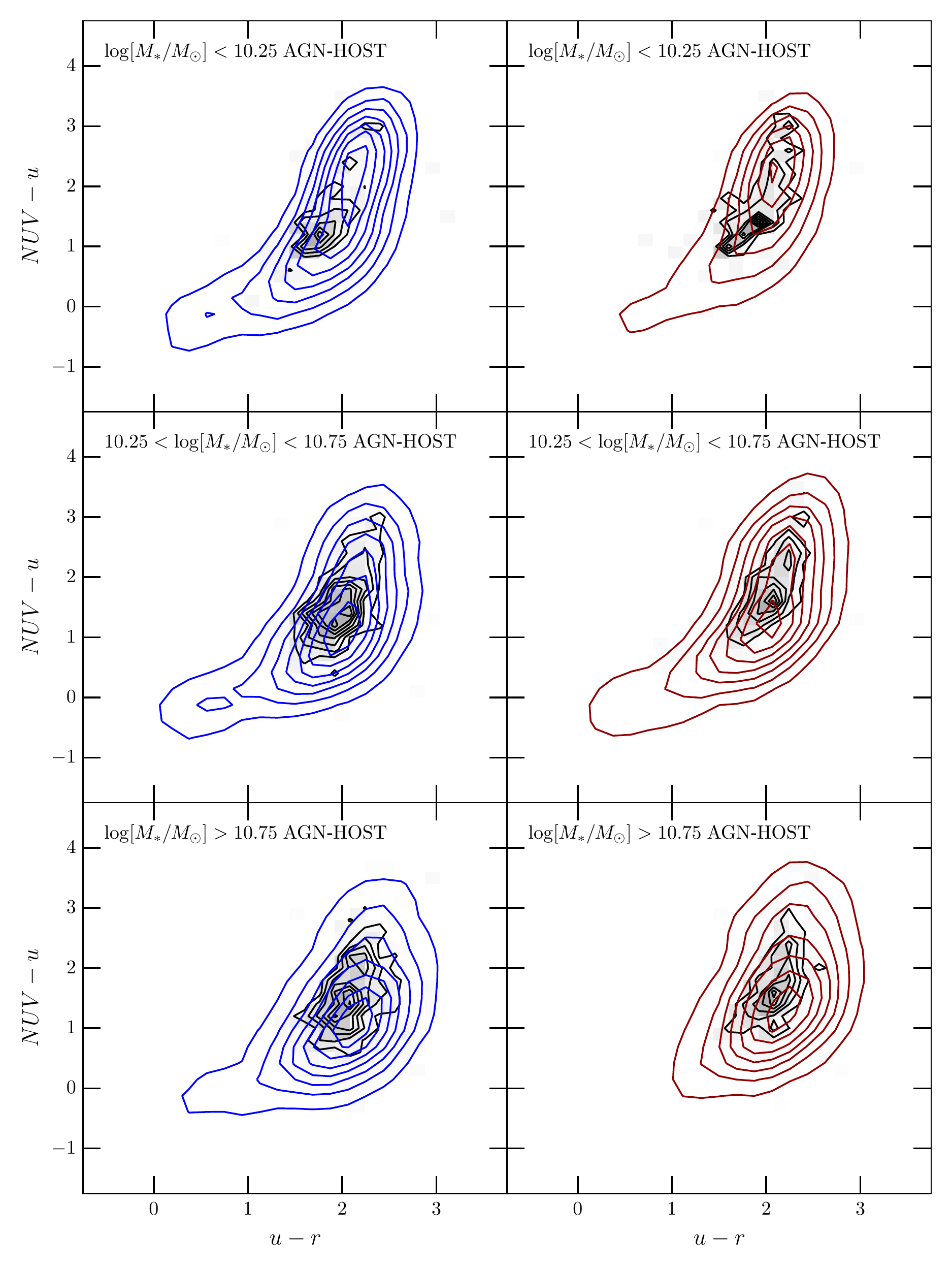}
\includegraphics[width=0.42\textwidth]{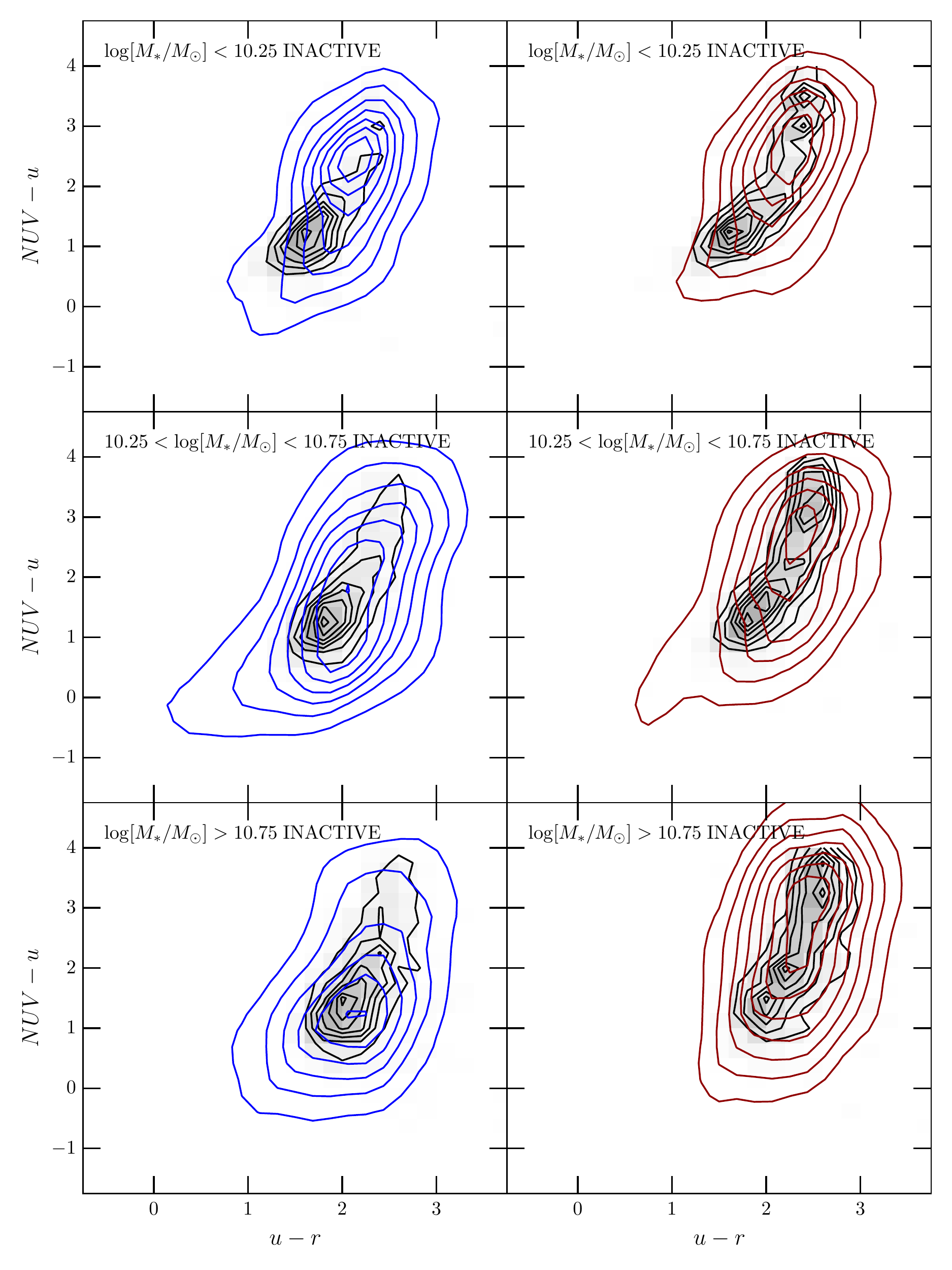}
\caption[8pt]{Optical-NUV colour-colour diagrams for the \textsc{agn-host} (top) and \textsc{inactive} (bottom) galaxies shown by the black contours, split into low mass (top), medium mass (middle) and high mass (bottom) galaxies weighted by $p_d$ (left) and $p_s$ (right). Kernel smoothing has been applied to the overlaid replica datasets, which are created by sampling from the \textbf{unweighted summed visualisation distributions} presented in Figures~\ref{time} and \ref{rate}. Gaussian random noise is also added to the inferred colours, with a mean and standard deviation of the errors on the observed colours of the respective sample. Contours are shown for samples taken from the disc (blue) and smooth weighted (red) summed visualisation distributions.
}
\label{replicapop}
\end{centering}
\end{figure}

We also varied the value of $N_s$ and found that increasing the number of samples drawn did not improve this fit for either the \textsc{agn-host} or \textsc{inactive} populations. Similarly increasing the number of components in the Gaussian mixture model did not immediately improve the accuracy of the fit.  We therefore concluded that this functional form of the population distribution was unsatisfactory. An extensive exploration of a wide variety of functional forms is necessary to ensure the correct conclusions are drawn from the data. Such an investigation is beyond the scope of this paper. 

The approach presented in the paper was motivated by the investigation increasing the number of samples, $N_s$ drawn from the posterior of each galaxy, k, until the point where all the samples were drawn. Instead of attempting to infer parameters to describe this distribution, as above, we presented the distribution itself.  The distributions produced by this visualisation method, shown in Figures~\ref{time} and \ref{rate}, reveal the complexity that the parent distribution must describe which, as we concluded earlier, cannot be effectively modelled.

We also tested whether this method is reasonable by producing replica datasets in optical-NUV colour space, as before, by drawing $1000$ $[t, \tau]$ values from the unweighted summed distributions presented in Figures~\ref{time} and \ref{rate}. These replica datasets are shown here in Figure~\ref{replicapop} in comparison to the observed colour-colour distributions of the \textsc{agn-host} and \textsc{inactive} samples. Comparing these replica colours in Figure~\ref{replicapop}, with those produced by drawing from the inferred hierarchical distributions, shown in Figure~\ref{replica}, we can see that they produce a more accurate match to the observed data for the majority of masses and morphologies. 

We therefore use this visualisation method to display the parent population distribution, rather than quoting inferred values to describe it.}

\end{document}